\documentclass[11pt]{article}
\usepackage{graphicx}

\begin{document}

\author{X. G. Wen$^{1,3}$ \& A. Zee$^{2,3}$ \\
%EndAName
$^{1}$ Department of Physics, Massachusetts Institute of Technology\\
Cambridge, Massachusetts 02139, USA \\
http://dao.mit.edu/\~\/wen\\
$^{2}$Institute for Theoretical Physics, University of California\\
Santa Barbara, California 93106, USA\\
zee@itp.ucsb.edu\\
$^3$ Center for Advanced Study, Tsinghua University\\
 Beijing 100084, P. R. China\\
}

\title{Gapless Fermions and Quantum Order}
\date{}
\maketitle

\begin{abstract}
Using 2D quantum spin-1/2 model as a concrete example, we studied the relation
between gapless fermionic excitations (spinons) and quantum orders in 
some spin liquid states. Using winding number, we find the
projective symmetry group that characterizes the quantum order directly
determines the pattern of Fermi points in the Brillouin zone. Thus quantum
orders provide an origin for gapless fermionic excitations.
\end{abstract}


\section{Introduction}

Gapless excitations are very rare in nature and in condensed matter systems.
Therefore, if we see a gapless excitation, we would like to ask why it exists.
One origin of gapless excitations is spontaneous symmetry breaking, which
gives Nambu-Goldstone bosons.\cite{N6080,G6154} However, spontaneous symmetry
breaking is not the only source of gapless excitations.

Recently, it was pointed out that the internal order of a generic quantum state 
cannot be characterized by broken symmetries and the associated order
parameter.  A new concept - quantum order - was introduced to describe those
internal orders.\cite{Wqos,Wqo,Wlight} It turns out that quantum orders
provide another origin for gapless excitations. The gapless excitations  from
quantum order are not always ordinary bosons, they can be gapless fermions or
gapless gauge bosons. 

To have a concrete description of quantum order, a mathematical object
-- projective symmetry group (PSG) -- was introduced.\cite{Wqos,Wqo}
PSG is a projective extension of symmetry group. The PSG description carries
more information than the symmetry group description since the same symmetry
group can have many different extensions.  Thus two quantum states with the
same symmetry can have different PSG's which correspond to different quantum
orders. The PSG description of quantum order generalize the symmetry group
description of classical orders.  It was by 
studying many examples of quantum ordered
states characterized by the PSG that we found that gapless excitations can
also be originated from and protected by quantum
orders.\cite{Wqos,Wqo,Wlight} 

%Quantum orders are characterized by projective symmetry groups
%(PSG).\cite{Wqos,Wqo} 
The relation between the gapless gauge fluctuations and PSG  has been
discussed in Ref.~\cite{Wqo,Wlight}. The relation is simple and
straightforward. The gauge group for the gapless gauge fluctuations in a
quantum state is found to be the so called invariant gauge group (IGG). IGG is
a (normal) subgroup of the PSG 
%(which characterizes the quantum order in that quantum state), 
such that $PSG/IGG$ gives rise to the symmetry group of the state.  This
result suggests that the gauge fields that govern all the interactions in
nature may have their origin in quantum order.\cite{Wlight}

The relation between the gapless fermions and projective symmetry groups are
more complicated.  However, through a study of most general fermion-hopping
Hamiltonian that is invariant under a PSG, one can show that the PSG can indeed
protect gapless fermions.\cite{Wqo} We would like to stress that those gapless
fermions are not chiral fermions and the gaplessness is not a result of chiral
symmetry. Thus quantum orders also provide an origin for gapless Dirac
fermions.\cite{Wlight} (Note that the model may contain no fermions at high
energies!)

In this paper, we are going to study the relation between PSG and gapless
fermion excitations using an index theorem. We will use the method discussed
in Ref. \cite{WZwind}. Through this study, we reveal a
more direct connection between PSG and gapless fermions. We can see clearly
how PSG protects gapless fermions.  The direct connection between PSG and
gapless fermions in the spin-1/2 model suggests that the relation between
gapless fermions and quantum orders may appear in much broader models.

It was pointed out that PSG alone cannot completely characterize quantum
orders.\cite{Wqos,Wqo} It is important to find other universal properties of
quantum states, so that we can use those properties to characterize quantum
orders in further details. Using PSG and index theorem, we find the
distributions of the crystal momenta of gapless fermions to have certain
topological invariance.  Thus the topology of those distributions can be used
to further characterize quantum orders.

As a concrete example, we will concentrate on a quantum spin-1/2 model on a 2D
square lattice.  Despite the absence of any solid proof, we believe that,
depending on the spin Hamiltonian, the 2D spin-1/2 model can have many different
ground states corresponding to different spin liquids.  Those different spin
liquids are described by different quantum orders and have different gapless
fermion excitations.  Those gapless fermions are called the spinons, which
carry spin-1/2 but no charge.  In this paper, we will study how the gapless
spinons are determined by the PSG that describes the quantum order. 

\section{Winding number and zero eigenvalues}

First let us introduce winding numbers.
Consider a $2n$ dimensional hermitian  matrix $H(\mathbf{k})$ that depends on
two parameters $\mathbf{k}=(k_x,k_y)$. We assume that $H(\mathbf{k})$
anticommutes with $\gamma^5$, where
\begin{equation}
 \gamma^5 =\left(\begin{array}{ll} I_{n\times n} & 0 \\
             0& -I_{n\times n} \end{array}\right)
\end{equation}
Thus 
\begin{equation}
 M(\mathbf{k})\equiv \frac{H(\mathbf{k})}{|\det(H(\mathbf{k}))|^{1/2n}}
\end{equation}
has a form
\begin{equation}
 M =\left(\begin{array}{ll} 0 & h^\dagger_{n\times n}  \\
         h_{n\times n}& 0 \end{array}\right)
\end{equation}
where $|\det(h)|=1$.  Now we see that if the phase of $\det(h)$ winds by
$2\pi$ around a loop in $\mathbf{k}$ space, then $\det(H(\mathbf{k}))$ must
vanish somewhere inside the loop and $H$ has zero eigenvalues at those points.
Thus we can determine the existence of zero eigenvalue for $H$ by calculating
the winding number around a loop.  To calculate the winding number, we
introduce
\begin{eqnarray}
 a_x &=& -i \partial_{k_x} \ln \det(h)  \nonumber\\
 a_y &=& -i \partial_{k_y} \ln \det(h)  
\end{eqnarray}
We find
\begin{eqnarray}
\label{axy}
 a_x &=& -i \frac{1}{2} \hbox{Tr} \gamma^5 M^{-1}\partial_{k_x} M  \nonumber\\
 a_y &=& -i \frac{1}{2} \hbox{Tr} \gamma^5 M^{-1}\partial_{k_y} M  
\end{eqnarray}
The winding number around a loop $C$ is defined as
\begin{equation}
 N= (2\pi)^{-1} \oint_C d\mathbf{k} \cdot \mathbf{a}
\end{equation}
which is quantized as an integer.  We note that $N$ does not change as we
deform the loop, as long as the deformation does not pass any point where
$\det(H(\mathbf{k}))=0$.  A non-zero winding number around a loop implies the
existence of zero eigenvalues of $H$ inside the loop.

We note that, for a $H(\mathbf{k})$, the winding number along a small loop
around a zero of $\det(H(\mathbf{k}))$ is $\pm 1$.  Due to the topological
property of the winding number, a zero of $\det(H(\mathbf{k}))$ cannot just
appear or disappear by itself as we deform the $H(\mathbf{k})$, as long as the
condition $[\gamma^5, H]=0$ is not violated.  Zeros can only be
created/annihilated in groups with vanishing total winding number.

\section{Gapless points in certain spin liquids}

In the slave-boson approach to 2D spin-1/2 system on a square
lattice,\cite{BZA8773,BA8880,AZH8845,DFM8826,Wsrvb} a spin liquid state is
described by the following mean-field Hamiltonian
\begin{eqnarray}
 H_{mean} &=& -\sum_{<\mathbf{i}\mathbf{j}>} \left( 
\psi^\dagger(\mathbf{i}) u_{\mathbf{i}\mathbf{j}} \psi(\mathbf{j}) + h.c. 
\right) 
 +\sum_{\mathbf{i}}  \psi^\dagger(\mathbf{i}) a_0^l \tau^l \psi(\mathbf{i})
\end{eqnarray}
where we have used the notations in Ref. \cite{Wqo}.  Here, $l=1,2,3$,
$\tau^l$ are the Pauli matrices, and $u_{\mathbf{i}\mathbf{j}}$ are complex
$2\times 2$ matrices that satisfy
\begin{equation}
 \hbox{Tr}( u_{\mathbf{i}\mathbf{j}})=\hbox{imaginary},\ \ \ \ \
 \hbox{Tr}( \tau^l u_{\mathbf{i}\mathbf{j}})=\hbox{real}
\end{equation}
The last condition ensure the spin liquid described by the ansatz
$u_{\mathbf{ij}} $ to have spin rotation symmetry.\cite{Wqo} One can see
clearly that $H_{mean}$ is invariant under the following $SU(2)$ gauge
transformation
\begin{eqnarray}
 \psi(\mathbf{i}) &\to&  G(\mathbf{i})~\psi(\mathbf{i})  \nonumber \\
 u_{\mathbf{i} \mathbf{j}} &\to& G(\mathbf{i})~u_{\mathbf{i} \mathbf{j}}~G^\dagger(\mathbf{j})
\end{eqnarray} 
where $G(\mathbf{i})\in SU(2)$.  We also note that the ansatz
$u_{\mathbf{i}\mathbf{j}}$ can be viewed as an operator which maps a fermion
wave function $\psi(\mathbf{i})$ to $\psi^\prime(\mathbf{i}) =
\sum_{\mathbf{j}} u_{\mathbf{i}\mathbf{j}} \psi(\mathbf{j})$. Such an operator
will be called the Hamiltonian whose eigenvalues determine the fermion
spectrum. The gapless fermions correspond to the zero eigenvalue of the
Hamiltonian.

The PSG of an ansatz $u_{{\mathbf{ij}}}$ is simply the symmetry group of the
ansatz. In this paper we are interested in the relation between the PSG and
the gapless fermions of the ansatz.
%We are interested in the symmetry of the ansatz  $u_{{\mathbf{ij}}}.$ 
Using the notation of \cite{Wqo}, we consider the set of transformations $U$
that acts on the lattice, consisting of translation, rotation, and so forth.
This induces a transformation on $u_{\mathbf{ij}}:$
$u_{\mathbf{ij}}\rightarrow U(u_{\mathbf{ij}})\equiv
u_{U(\mathbf{i)}U\mathbf{(j)}}.$ When we translate, for example, we may get a
$u_{\mathbf{ij}}$ that looks different from the $u_{\mathbf{ij}}$ we started
with, but the two $u_{\mathbf{ij}}$'s may in fact be related by a gauge
transformation. For each $U$ we define a gauge transformation
$G_{U}(\mathbf{i})\epsilon SU(2)$ which induces a transformation on
$u_{\mathbf{ij}}:u_{\mathbf{ij}}\rightarrow G_{U}(u_{\mathbf{ij}})\equiv
G_{U}(\mathbf{i})u_{\mathbf{ij}}G_{U}^{\dagger}(\mathbf{j}).$ 
A given ansatz has a symmetry if and only if it is invariant under the
corresponding transformation $U $ followed by a proper gauge transformation
$G_{U}$: \begin{equation} G_{U}U(u_{\mathbf{ij}})=u_{\mathbf{ij}}
\end{equation}

The group generated by the set of all combined transformations $G_{U}U$ that
leave the ansatz invariant is called the $PSG$. The $IGG$ is defined as a
subgroup of the PSG generated by $G_{I},$ that is, by the $G_{U}$ for $ U=I$.
In other words, the $IGG$ consists of pure gauge transformations that leave
the form of $u_{\mathbf{ij}}$ invariant. IGG is a normal subgroup of PSG which
allows us to define the coset $SG\equiv PSG/IGG$. $SG$ is nothing but the
symmetry group generated by all the $U$'s.

\subsection{Gapless fermions in Z2A0013 spin liquids}

Now we are ready to study the direct relation between the PSG and gapless
fermions.  Let us first discuss a simple symmetric $Z_2$-linear spin liquid
characterized by a PSG labeled by Z2A0013.\cite{Wqo} The Z2A0013 spin liquid
is described by a large class of ansatz that are invariant under the Z2A0013
PSG \cite{Wqo}. One simple example of those ansatz is given by
\begin{eqnarray}
\label{Z2A0013}
 u_{\mathbf{i},\mathbf{i}+\hat{\mathbf{x}}} &=& (\chi \tau^1 - \eta\tau^2) 
\nonumber\\
 u_{\mathbf{i},\mathbf{i}+\hat{\mathbf{y}}} &=& (\chi \tau^1 + \eta\tau^2)  
\nonumber\\
 u_{\mathbf{i},\mathbf{i}+\hat{\mathbf{x}}+\hat{\mathbf{y}}} &=& 
+ \chi_1 \tau^1 \nonumber\\
 u_{\mathbf{i},\mathbf{i}-\hat{\mathbf{x}}+\hat{\mathbf{y}}} &=& 
+ \chi_1 \tau^1 \nonumber\\
 u_{\mathbf{i},\mathbf{i}+2\hat{\mathbf{x}}} &=& 
\chi_2\tau^1 + \eta_1 \tau^2 \nonumber\\
 u_{\mathbf{i},\mathbf{i}+2\hat{\mathbf{y}}} &=& 
\chi_2\tau^1 - \eta_1 \tau^2 
\nonumber\\
 a^1_0 & \neq & 0,\ \ \ \ \ a^{2,3}_0=0
\end{eqnarray}
More general ansatz can be found in Ref. \cite{Wqo}.  The ansatz is invariant
under the following combined transformations $\{ G_0$, $ G_xT_x$, $ G_yT_y$, $
G_{P_x}P_x$, $G_{P_y}P_y$, $ G_{P_{xy}}P_{xy}$, $ G_TT\}$ that act on the
fermion wave functions.  They are formed by a symmetry transformation followed
by a gauge transformation. The symmetry transformations are two translations
$T_x$: $\psi(i_x,i_y)\to\psi(i_x-1,i_y)$ and $T_y$:
$\psi(i_x,i_y)\to\psi(i_x,i_y-1)$, three parities $P_x$:
$\psi(i_x,i_y)\to\psi(-i_x,i_y)$, $P_y$: $\psi(i_x,i_y)\to\psi(i_x,-i_y)$, and
$P_{xy}$: $\psi(i_x,i_y)\to\psi(i_y,i_x)$, and a time reversal transformation
$T$: $T(u_{\mathbf{i}\mathbf{j}}) = -u_{\mathbf{i}\mathbf{j}}$.\cite{Wqo} The
corresponding gauge transformations are given by
\begin{eqnarray}
 G_x(\mathbf{i}) &=& \tau^0,\ \ \ \ \ G_y(\mathbf{i}) = \tau^0,  \nonumber\\
 G_{P_x}(\mathbf{i}) &=& \tau^0,\ \ \ \ \ G_{P_y}(\mathbf{i}) = \tau^0,  \nonumber\\
 G_{P_{xy}}(\mathbf{i}) &=& i\tau^1,\ \ \ \ \ G_T(\mathbf{i}) = i\tau^3, \nonumber\\
 G_0(\mathbf{i}) &=& -\tau^0, 
\end{eqnarray}
where $\tau^0$ is the $2\times 2$ identity matrix. The above defines the
Z2A0013 PSG.
% defined as the
%invariant group of the ansatz,\cite{Wqos,Wqo} is found to be generated by the
%above combined transformations. 
We note that the combined transformation, such as $G_xT_x$ is a linear
operator that act on the fermion wave functions $\psi$.  From the operator
point of view, the invariance of the ansatz implies and requires that all the
elements in the PSG commute with the Hamiltonian $u_{\mathbf{i}\mathbf{j}}$,
except $G_T$ which anticommutes with $u_{\mathbf{i}\mathbf{j}}$.  The IGG, a
(normal) subgroup of the PSG, is formed by the pure gauge transformations in
the PSG. We find that IGG is generated by $G_0$ and $IGG=Z_2$. Therefore, the
above ansatz describes a $Z_2$ spin liquids.

In the momentum space, the mean-field Hamiltonian becomes
\begin{eqnarray}
\label{Hk0013}
 H(\mathbf{k})=\epsilon(\mathbf{k})\tau^1+\Delta(\mathbf{k})\tau^2
\end{eqnarray}
The spinon spectrum is given by
\begin{equation}
 E_\pm(\mathbf{k}) =\pm \sqrt{\epsilon^2(\mathbf{k}) + \Delta^2(\mathbf{k})}
\end{equation}
where
\begin{eqnarray}
 \epsilon &=&  a^1_0+ 2\chi( \cos(k_x)+\cos(k_y)) \nonumber\\
  &&\ \ + 2\chi_1( \cos(k_x+k_y)+\cos(k_x-k_y))  \nonumber\\
  &&\ \ + 2\chi_2( \cos(2k_x)+\cos(2k_y))  \nonumber\\
 \Delta &=& 2\eta(-\cos(k_x)+\cos(k_y))  \nonumber\\
  &&\ \ + 2\eta_1( \cos(2k_x)-\cos(2k_y)) .
\end{eqnarray}
%The spinon eigenstate with energy $E_\pm(\mathbf{k})$ has a form
%\begin{equation}
% \Phi_{\mathbf{k}} = \left(\begin{array} u(\mathbf{k}) \\ v(\mathbf{k}) \end{array}\right)
%\end{equation}
We see that $\tau^3$ anticommutes with $H(\mathbf{k})$.  We note that
$G_T=i\tau^3$ and the Z2A0013 ansatz always change sign under the time reversal
transformation $G_T$.  Thus, if we choose $\gamma^5=\tau^3$, $\gamma^5$ will
anticommute with $H(\mathbf{k})$ for any Z2A0013 ansatz.  In this case, we can
define $\mathbf{a} = (a_x,a_y)$ via Eq.~\ref{axy}, which allow us to calculate
topological invariant winding numbers and determine the zeros of
$H(\mathbf{k})$ for Z2A0013 ansatz. 

Now let us consider the symmetry of $H(\mathbf{k})$ as required by the
invariance of the ansatz under the Z2A0013 PSG.  By considering the action of
$G_{P_x}T_{P_x}$ and $G_{P_y}T_{P_y}$ on the $\mathbf{k}$-space Hamiltonian
$H(\mathbf{k})$, we find that the invariance of the ansatz under
$G_{P_x}T_{P_x}$ leads to
\begin{eqnarray}
U_{P_x} H(k_x,k_y) U^\dagger_{P_x}  &=& H(-k_x,k_y),  \nonumber\\
U_{P_x} &=& \tau^0
\end{eqnarray}
and the invariance of the ansatz under $G_{P_y}T_{P_y}$ leads to
\begin{eqnarray}
U_{P_y} H(k_x,k_y) U^\dagger_{P_y}  &=& H(k_x,-k_y),  \nonumber\\
U_{P_y} &=& \tau^0
\end{eqnarray}
The invariance of the ansatz under $G_{P_{xy}}T_{P_{xy}}$ gives us a little
less trivial result
\begin{eqnarray}
U_{P_{xy}} H(k_x,k_y) U^\dagger_{P_{xy}}  &=& H(k_y,k_x),  \nonumber\\
U_{P_{xy}} &=& i\tau^1
\end{eqnarray}
We would like to stress that the above results are valid for any Z2A0013
ansatz. Together with $\{\tau^3, H\} = 0$, we find that
the Hamiltonian $H(\mathbf{k})$ for Z2A0013 ansatz must take the form
Eq. (\ref{Hk0013}) with
\begin{eqnarray}
\label{sym0013}
&& \epsilon(k_x,k_y)=\epsilon(-k_x,k_y)=\epsilon(k_x,-k_y)=\epsilon(k_y,k_x)  
\nonumber\\
&& \Delta(k_x,k_y)=\Delta(-k_x,k_y)=\Delta(k_x,-k_y)=-\Delta(k_y,k_x)  
\end{eqnarray} 
In fact, Eqs. (\ref{Hk0013}) and (\ref{sym0013}) define the most general
Z2A0013 ansatz.

Now, let us study the symmetries of $\mathbf{a}$.  
Under $G_{P_x}P_x$ and $G_{P_y}P_y$, we have
\begin{equation}
 M(k_x,k_y)=M(-k_x,k_y), \ \ 
 M(k_x,k_y)=M(k_x,-k_y).
\end{equation}
Thus
\begin{eqnarray}
 a_x(k_x,k_y) &=&  -a_x(-k_x,k_y), \ \ \ \ \ a_y(k_x,k_y) =  a_y(-k_x,k_y),
\nonumber\\
 a_x(k_x,k_y) &=&  a_x(k_x,-k_y), \ \ \ \ \ a_y(k_x,k_y) = - a_y(k_x,-k_y).
\end{eqnarray}
Since $G_T$ commute with $G_{P_x}P_x$ and $G_{P_y}P_y$, the actions
of $G_{P_x}P_x$ and $G_{P_y}P_y$ on $\mathbf{a}$ are the same as the actions
of $P_x$ and $P_y$ respectively.

\begin{figure}
\centerline{
\includegraphics[height=1.5in]{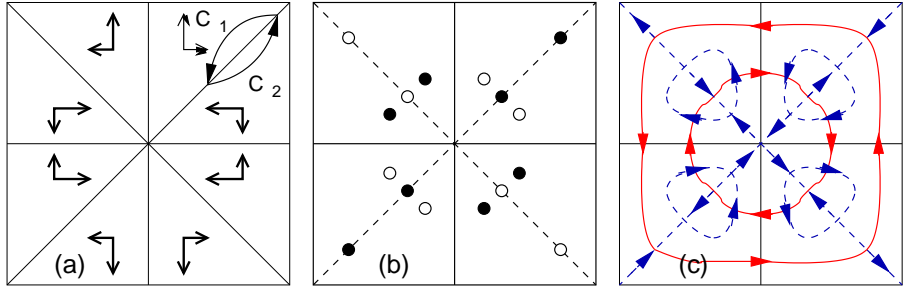}
}
\caption{
(a) The symmetry of $\mathbf{a}$ in the Brillouin zone. The two arrows
represent $a_x \hat{\mathbf{x}}$ and $a_y \hat{\mathbf{y}}$.
(b) A typical distribution of the zeros of $\det(H(\mathbf{k}))$. A solid dot
represents a zero with winding number $+1$ and an empty dot represents a zero
with winding number $-1$.   The solid-lines and the dash-lines (including the
ones on zone boundary) are lines of
mirror symmetry, whose precise meaning is given in Fig. \ref{trnT1}.
(c) The curves $\epsilon=0$ (solid lines) and $\Delta = 0$ (dash lines). The
direction of the curves is defined such that the right-hand-side of the curve
has a positive $\epsilon$ (or $\Delta$).
}
\label{fig1}
\end{figure}

Next, we consider the
action of $G_{P_{xy}}P_{xy}$. From the relation between
$H(\mathbf{k})$ and $M(\mathbf{k})$, we have
\begin{equation}
 \tau^1 M(k_x, k_y) \tau^1 = M(k_y, k_x) .
\end{equation}
From the definition
\begin{eqnarray}
&&a_x(k_x,k_y)\nonumber\\
&=&-i\frac{1}{2}\hbox{Tr}[\tau^3M^{-1}(\mathbf{k}) \partial_{k_x} M(\mathbf{k})]\nonumber\\
&=&-i\frac{1}{2}\hbox{Tr}[\tau^3M^{-1}(\mathbf{k}) \partial_{k_x}\tau^1 M(k_y,k_x)\tau^1]\nonumber\\
&=&i\frac{1}{2}\hbox{Tr}[\tau^3M^{-1}(k_y,k_x)\partial_{k_x}
      M(k_y,k_x)]\nonumber\\
&=&-a_y(k_y,k_x) .
\end{eqnarray}
We note that $P_{xy}$ changes $(a_x(k_x,k_y), a_y(k_x,k_y))\to (a_y(k_y,k_x),
a_x(k_y,k_x))$. From the Z2A0013 PSG, we see that $G_{P_{xy}}P_{xy}$ and $G_T$
(which is $\gamma^5$) anticommute.  Therefore $G_{P_{xy}}P_{xy}$ changes
$(a_x(k_x,k_y), a_y(k_x,k_y))\to -(a_y(k_y,k_x), a_x(k_y,k_x))$, due to the
$\gamma^5$ in the definition of $\mathbf{ a}$. The invariance of the ansatz
under $G_{P_{xy}}P_{xy}$ implies
\begin{equation}
a_x(k_x,k_y) = -a_y(k_y,k_x),\ \ \ \ \
a_y(k_x,k_y) = -a_x(k_y,k_x)
\end{equation}
We see that the symmetry of $\mathbf{a}$ is directly related the algebraic
property of the corresponding PSG.
The symmetry of $\mathbf{a}$ is summarized in Fig. \ref{fig1}a

From the symmetry, we see that the winding numbers vanishes for small square
loops centered along the lines $[(0,0), (\pm\pi, 0)]$, $[(0,0), (0, \pm\pi)]$,
$[(\pi,\pm\pi), (\pi, 0)]$,  and $[(\pm\pi,\pi), (0, \pi)]$.  Therefore, a
generic Z2A0013 ansatz does not have any gapless fermions along those lines.
Now consider a loop $C$ formed by the two segment $C_1$ and $C_2$ in Fig.
\ref{fig1}a. $C_1$ and $C_2$ are related through $P_{xy}$. We find that
$\int_{C_1} d\mathbf{k}\cdot \mathbf{a} = \int_{C_2} d\mathbf{k}\cdot
\mathbf{a} $.  Thus, $2(2\pi)^{-1}\int_{C_1} d\mathbf{k}\cdot \mathbf{a} $ is
quantized as an integer, which is the algebraic sum of the numbers of zeros
enclosed by the loop formed by $C_1$ and $C_2$.  Thus, when
$2(2\pi)^{-1}\int_{C_1} d\mathbf{k}\cdot \mathbf{a} $ is an odd integer,
$\det(H(\mathbf{ k}))$ must vanish at least at one point on the line
$((0,0),(\pi,\pi))$.

Noticing that a zero of $\det(H(\mathbf{k}))$ typically has $\pm 1$ winding
number, when combined with the symmetry of $\mathbf{a}$, we find a typical
distribution of the zero of $\det(H(\mathbf{k}))$ may look like Fig.
\ref{fig1}b.  The pattern of the winding number is determined by noting that
$G_{P_x}P_x$ map a zero at $(k_x,k_y)$ with winding number $n$ to a zero at
$(-k_x,k_y)$ with winding number $-n$.  Thus action of $G_{P_x}P_x$ is the
same as the action of ordinary parity transformation $P_x$ which change the
sign of winding number. This is because $G_T$ and $G_{P_x}P_x$ commute.
Similarly,  $G_{P_y}P_y$ map a zero at $(k_x,k_y)$ with winding number $n$ to
a zero at $(k_x,-k_y)$ with winding number $-n$.  However, $G_{P_{xy}}P_{xy}$
map a zero at $(k_x,k_y)$ with winding number $n$ to a zero at $(k_y,k_x)$
with winding number $n$. It is strange to see that the $P_{xy}$ parity
transformation does change the sign of winding number.  This is because $G_T$
and $G_{P_{xy}}P_{xy}$ anticommute.  We would like to stress that \emph{the
pattern of the winding numbers and the positions of the zeros are directly
determined from the algebraic property of the PSG.} 

For the Z2A0013 state, the pattern of zeros can also be determined from Eq.
(\ref{sym0013}). Zeros appear at the crossing point of the $\epsilon = 0$ and
$\Delta = 0$ curves, as shown in Fig. \ref{fig1}c.

Physically, each zero correspond to a massless two-component Dirac fermion in
1+2 dimensions.  We see that in general the Z2A0013 state can have $4n$
two-component Dirac fermions. In particular, Z2A0013 spin liquids can have no
gapless fermions if $a_0^1$ is large enough.

The topological property of the winding number and the symmetry of
$\mathbf{a}$ imply that the topology of the zeros (TOZ) is a universal
property of the quantum state. (It was shown that the topology of Fermi
surfaces determine the quantum orders in free fermion systems.\cite{Wqos} Here
the topology of zeros is a special case of topology of Fermi surfaces.) TOZ is
not changed by any small deformations of the ansatz, as long as the PSG is not
changed.  We see that the quantum orders in the spin liquid are characterized
by (at least) two universal properties: PSG and TOZ.\cite{Wqos} PSG alone
cannot completely characterize the quantum orders.  For Z2A0013 spin liquids,
the TOZ, by definition, is a collection of the following data: (a) a pattern
of $+1$ and $-1$ zeros along the line $((0,0),(\pi,\pi))$, 
(b) $(N_+, N_-)$, the number of $(+1, -1)$ zeros
\emph{inside} the triangle $((0,0),(\pi,\pi), (0,\pi))$ (not including the
zeros on the sides and corners).  (a) is a universal property because the
zeros along the line $((0,0),(\pi,\pi))$ cannot move off the line by itself
due to the symmetry of $H(\mathbf{k})$.

\begin{figure}
\centerline{
\includegraphics[height=1.in]{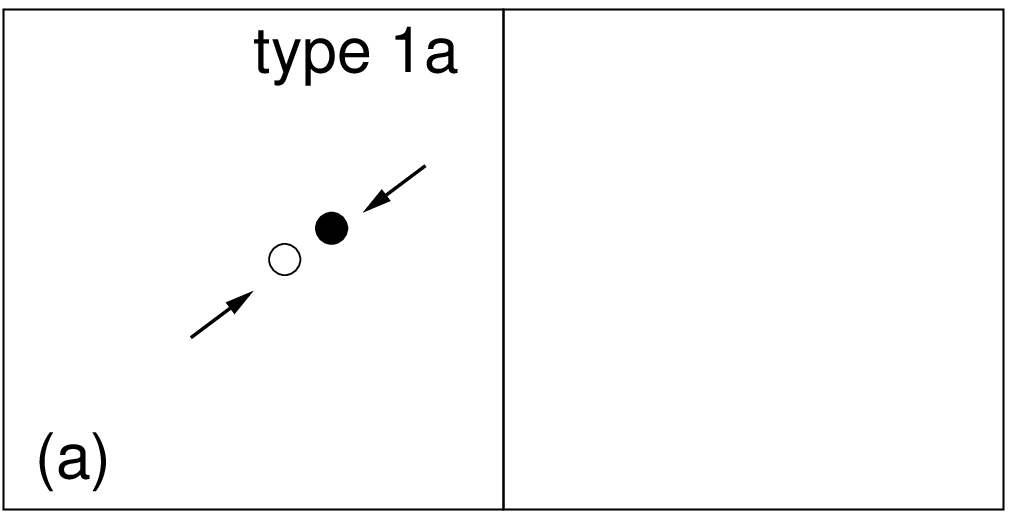}
\includegraphics[height=1.in]{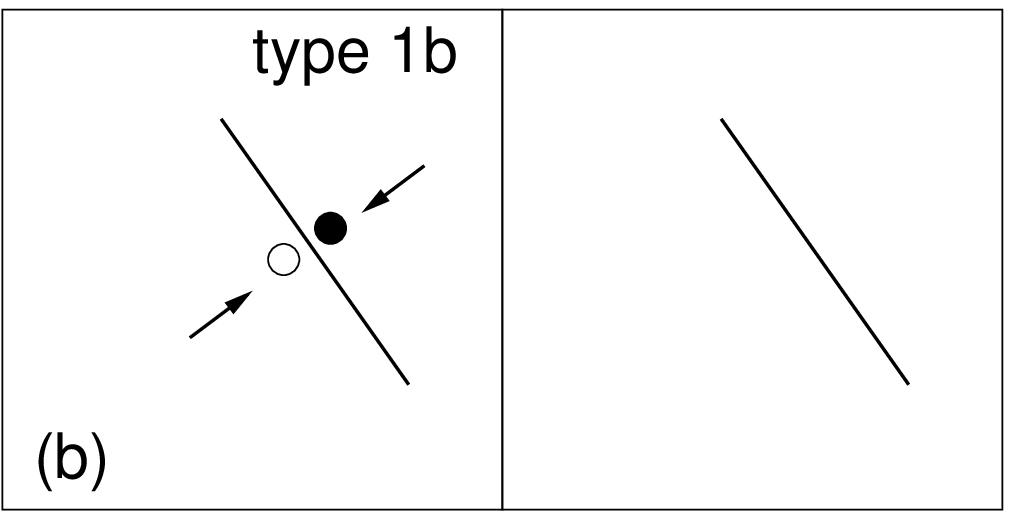}
}
\centerline{
\includegraphics[height=1.in]{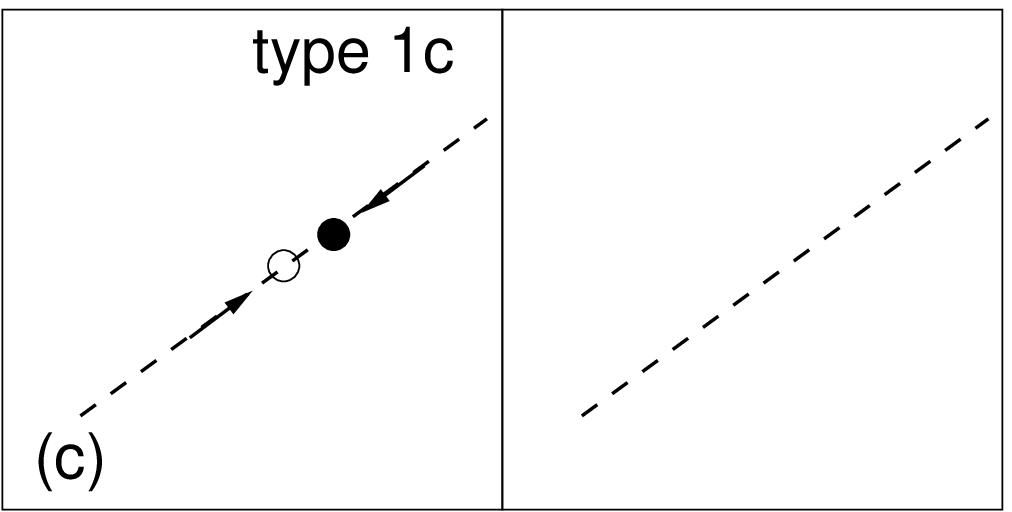}
\includegraphics[height=1.in]{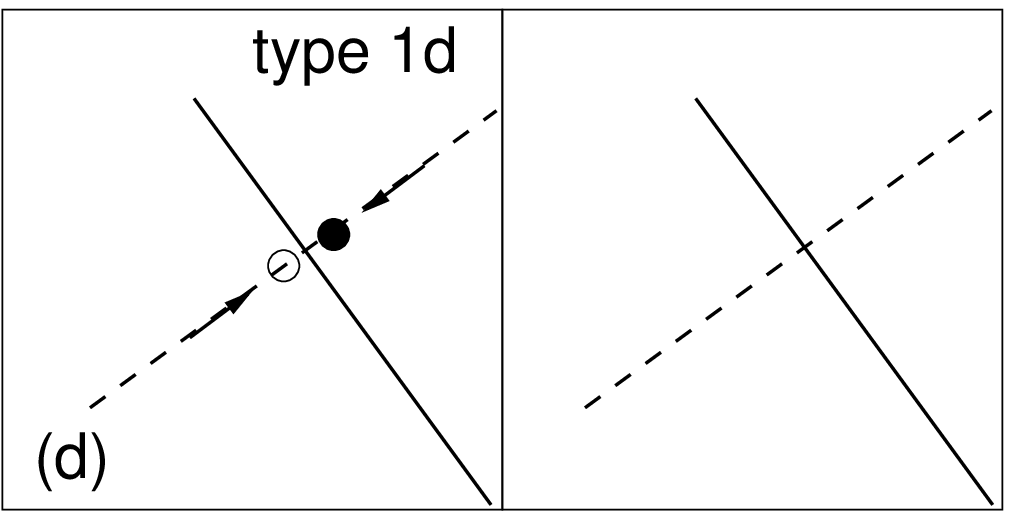}
}
\caption{
Type 1 transitions in different symmetry environments. 
Those transitions are named type 1a, type 1b, type 1c, and type 1d.
For each case the left and the right panels describe the nodes before and
after the transition respectively. Each case can be ``time reverse'' of cause.
The solid-line and
dash-line represent line of mirror symmetry. Zeros on the two sides of
solid-line have opposite winding numbers and zeros on the two sides of
dash-line have the same winding number.
Note, the square does not represent a Brillouin zone.
}
\label{trnT1}
\end{figure}

\begin{figure}
\centerline{
\includegraphics[height=1.in]{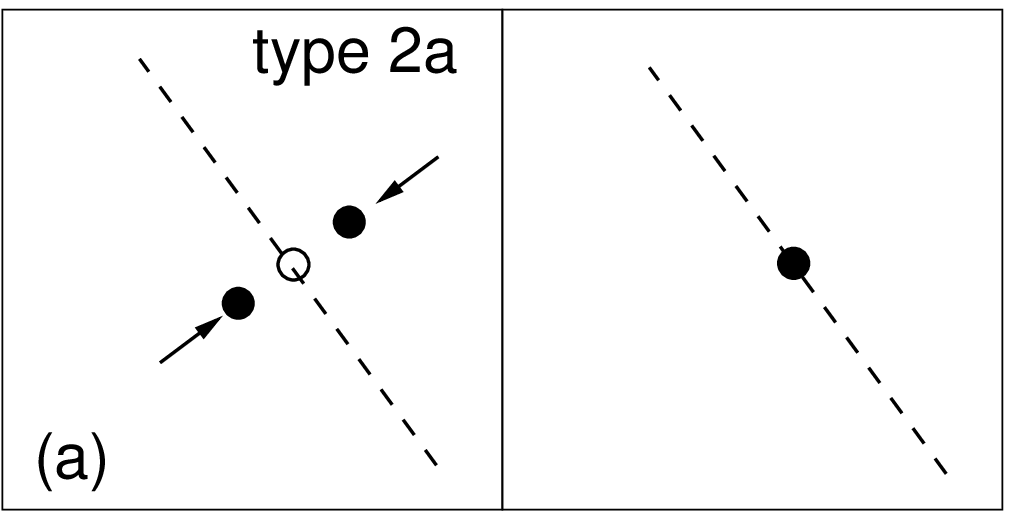}
\includegraphics[height=1.in]{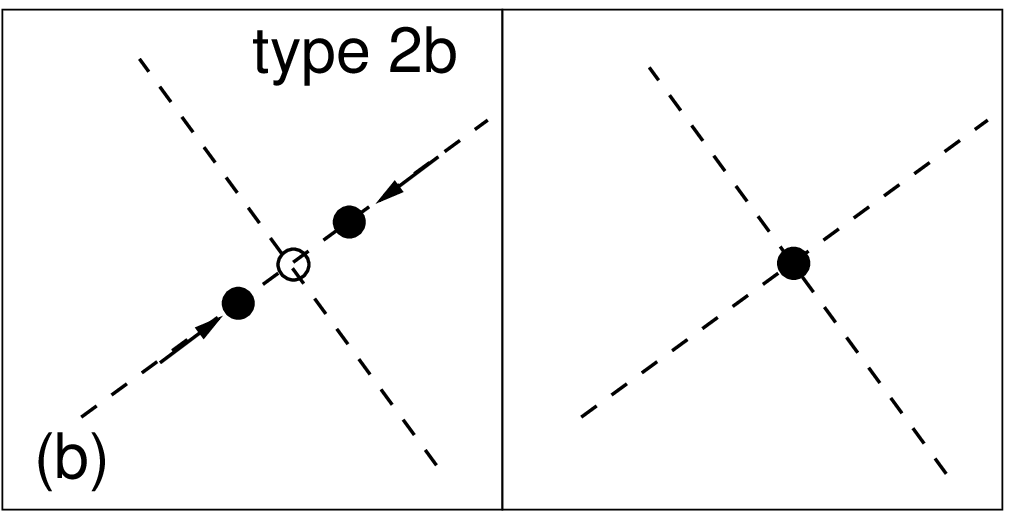}
}
\caption{
Type 2 transitions in different symmetry environments. 
Those transitions are named type 2a and type 2b.
}
\label{trnT2}
\end{figure}

\begin{figure}
\centerline{
\includegraphics[height=1.in]{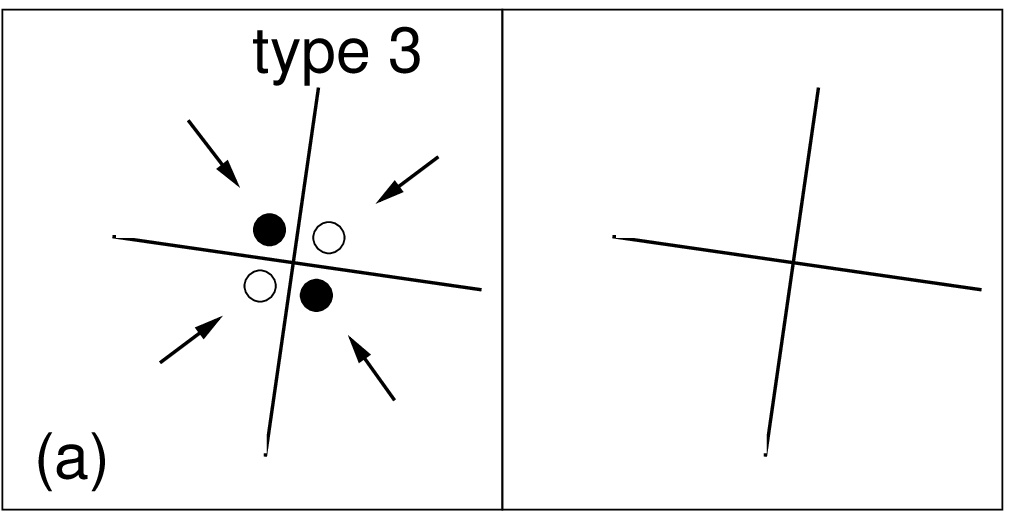}
\includegraphics[height=1.in]{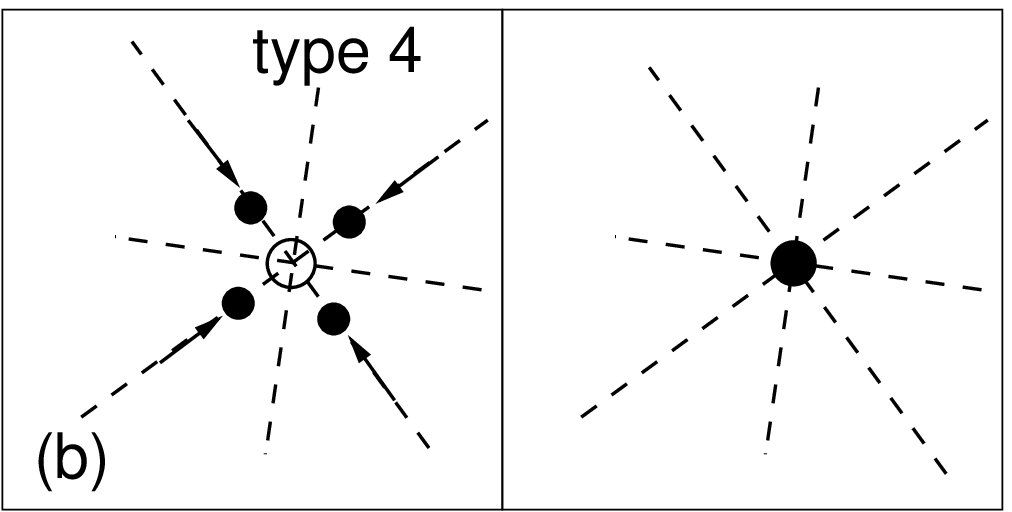}
}
\caption{
(a) Type 3 transition.
(b) Type 4 transition.
The winding number of large dots is twice as large as that of small dots.
}
\label{trnT34}
\end{figure}

Now let us consider the transitions between quantum orders with a fixed PSG
Z2A0013. Those transitions involve a change in TOZ. However, the symmetry of
the state is not changed. 

First we would like to have a general discussion of the transitions.  By
examining the possible ways in which TOZ can change that are consistent with
the symmetry of $H(\mathbf{k})$, we find there are several classes of
transitions.  Some of them are described in Figs.  \ref{trnT1},  \ref{trnT2},
and \ref{trnT34}.  In the type 1 transition, a pair of zeros with opposite
winding number is created (or annihilated). A type 2 transition involve three
zeros, two of them have the same winding number and the third one has the
opposite winding number.  In the transition, three zeros combine into one
zero.  In type 3 transition, two pairs of zeros with opposite winding number
are created (or annihilated). 
%In type 4 transition four zeros of the same winding number passing through
%each other.  Type 5 transition has two zeros carrying positive one unit of
%winding numbers colliding with a zero with negative one unit of winding
%numbers and changing it into a zero with positive one unit of winding
%numbers.  
In type 4 transition, four zeros carrying a unit of winding numbers changes a
zero with negative 2 units of winding numbers into a zero with positive 2
units of winding numbers.

Applying the above results to the Z2A0013 state,
we find that the possible quantum phase
transitions (caused by tuning one parameter) can be divided, at least, 
into the follow 5
classes. Because of the symmetry, we only need to describe the motion of zeros
near a triangle $((0,0), (\pi,0),(\pi,\pi))$.\\
(A) A type 1b transition on $((0,0), (\pi, 0))$ or $((\pi,0), (\pi, \pi))$
sides of the triangle.  \\
(B) A type 1c transition on on the line $((0,0), (\pi,\pi))$. \\
(C) A type 3 transition at $\mathbf{k}=(0,0)$ or $(\pi,\pi)$.\\
(D) A type 2a transition on the line $((0,0), (\pi,\pi))$ .  \\
(E) A type 1a transition \emph{inside} the triangle $((0,0),(\pi/2,\pi/2),
(0,\pi/2))$.

\subsection{Gapless fermions in Z2Azz13 spin liquids}

Next we consider the Z2Azz13 spin liquid state. One simple example
of the Z2Azz13 ansatz is given by:\cite{Wqo}
\begin{eqnarray}
\label{Z2Azz13}
u_{\mathbf{i},\mathbf{i}+\hat{\mathbf{x}}} &=& \chi \tau^1 - \eta\tau^2 \nonumber\\
u_{\mathbf{i},\mathbf{i}+\hat{\mathbf{y}}} &=& \chi \tau^1 + \eta\tau^2 \nonumber\\
 u_{\mathbf{i},\mathbf{i}+\hat{\mathbf{x}}+\hat{\mathbf{y}}} &=& - \chi_1 \tau^1 \nonumber\\
 u_{\mathbf{i},\mathbf{i}-\hat{\mathbf{x}}+\hat{\mathbf{y}}} &=& + \chi_1 \tau^1 \nonumber\\
 u_{\mathbf{i},\mathbf{i}+2\hat{\mathbf{x}}} &=& 
 u_{\mathbf{i},\mathbf{i}+2\hat{\mathbf{y}}} = 0
 \nonumber\\
 a^{1,2,3}_0 & = &0
\end{eqnarray}
The Z2Azz13 PSG is generated by
\begin{eqnarray}
 G_x(\mathbf{i}) &=& \tau^0,\ \ \ \ \ G_y(\mathbf{i}) = \tau^0,  \nonumber\\
 G_{P_x}(\mathbf{i}) &=& i(-)^{i_x+i_y}\tau^3,\ \ \ \ \ 
 G_{P_y}(\mathbf{i}) = i(-)^{i_x+i_y}\tau^3,  \nonumber\\
 G_{P_{xy}}(\mathbf{i}) &=& i\tau^1,\ \ \ \ \ G_T(\mathbf{i}) = i\tau^3, 
\nonumber\\
 G_0(\mathbf{i}) &=& -\tau^0, 
\end{eqnarray}
In the momentum space, we have 
\begin{eqnarray}
\label{Hkzz13}
 H(\mathbf{k})=\epsilon(\mathbf{k})\tau^1+\Delta(\mathbf{k})\tau^2
\end{eqnarray}
where
\begin{eqnarray}
 \epsilon &=& 2\chi(\cos(k_x)+\cos(k_y)) \nonumber\\
  &&\ + 2\chi_1(\cos(k_x+k_y)-\cos(k_x-k_y)) 
      \nonumber\\
 \Delta &=& 2\eta(-\cos(k_x)+\cos(k_y))  
\end{eqnarray}
We see that the time reversal transformation $G_T$ gives us again
$\gamma^5=\tau^3$ which anticommutes with $H(\mathbf{k})$ for any Z2Azz13
ansatz.  We can use $\mathbf{a} = (a_x,a_y)$ to calculate the topological
invariant winding numbers and determine the zeros of $H(\mathbf{k})$ for any
Z2Azz13 ansatz. 

Now let us consider the symmetry of $H(\mathbf{k})$ as required by the the
Z2Azz13 PSG.  We find that the invariance of the ansatz under $G_{P_x}T_{P_x}$
leads to
\begin{eqnarray}
U_{P_x} H(k_x,k_y) U^\dagger_{P_x}  &=& H(-k_x+\pi,k_y+\pi),  \nonumber\\
U_{P_x} &=& i\tau^3
\end{eqnarray}
and the invariance of the ansatz under $G_{P_y}T_{P_y}$ leads to
\begin{eqnarray}
U_{P_y} H(k_x,k_y) U^\dagger_{P_y}  &=& H(k_x+\pi,-k_y+\pi),  \nonumber\\
U_{P_x} &=& i\tau^3
\end{eqnarray}
The invariance of the ansatz under $G_{P_{xy}}T_{P_{xy}}$ gives us 
\begin{eqnarray}
U_{P_{xy}} H(k_x,k_y) U^\dagger_{P_{xy}}  &=& H(k_y,k_x),  \nonumber\\
U_{P_x} &=& i\tau^1
\end{eqnarray}
Thus, the Hamiltonian $H(\mathbf{k})$ for Z2Azz13 ansatz must take the
form Eq. (\ref{Hkzz13}) with
\begin{eqnarray}
\label{symzz13}
&& \epsilon(k_x,k_y)=-\epsilon(-k_x+\pi,k_y+\pi)
                    =-\epsilon(k_x+\pi,-k_y+\pi)
                    = \epsilon(k_y,k_x)  
\\
&& \Delta(k_x,k_y)=-\Delta(-k_x+\pi,k_y+\pi)
                  =-\Delta(k_x+\pi,-k_y+\pi)
                  =-\Delta(k_y,k_x)   \nonumber 
\end{eqnarray} 
In fact, Eqs. (\ref{Hkzz13}) and (\ref{symzz13}) define the most general
Z2Azz13 ansatz.

\begin{figure}
\centerline{
\includegraphics[height=1.5in]{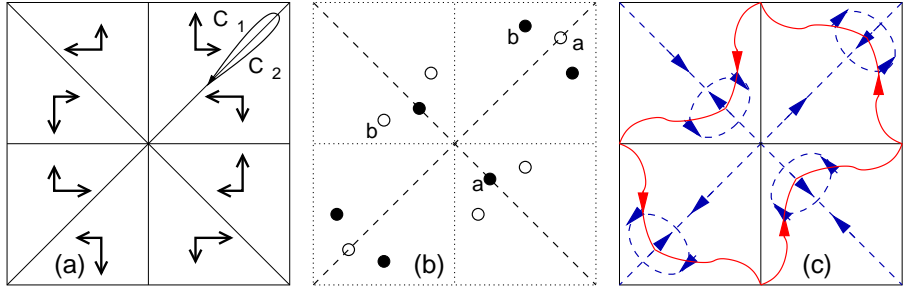}
}
\caption{
(a) The symmetry of $\mathbf{a}$ in the Brillouin zone for the Z2Azz13 state. 
The two arrows represent $a_x \hat{\mathbf{x}}$ and $a_y \hat{\mathbf{y}}$.
(b) A typical distribution of the zeros of $\det(H(\mathbf{k}))$. A solid dot represents a
zero with winding number $+1$ and an empty dot represents a
zero with winding number $-1$. 
The two dots marked by ``a'' are related by the
$G_{P_x}P_x$ transformation which changes $(k_x, k_y)\to (-k_x+\pi, k_y+\pi)$.
The two dots marked by ``b'' are related by the
$G_{P_y}P_y$ transformation which changes $(k_x, k_y)\to (k_x+\pi, -k_y+\pi)$.
(c) The curves $\epsilon=0$ (solid lines) and $\Delta = 0$ (dash lines). 
%The direction of the curves allows us to define the righ-hand-side of the curve
%where $\epsilon$ (or $\Delta$) is positive.
}
\label{fig2}
\end{figure}

Next, let us study the symmetries of $\mathbf{a}$.
From the Z2Azz13 PSG, we see that
$G_{P_{xy}}P_{xy}$ and $G_T$ anticommute.
The invariance under $G_{P_{xy}}P_{xy}$ implies
\begin{eqnarray}
a_x(k_x,k_y) = -a_y(k_y,k_x),\ \ \ \ \
a_y(k_x,k_y) = -a_x(k_y,k_x)
\end{eqnarray} 
We also see that
$G_{P_x}P_x$ and $G_T$ commute and $G_{P_x}P_x$ changes $(k_x,k_y)\to
(-k_x+\pi,k_y+\pi)$.
Therefore, the invariance under $G_{P_x}P_x$ leads to
\begin{eqnarray}
a_x(k_x,k_y) &=& -a_x(-k_x+\pi,k_y+\pi), \nonumber\\
a_y(k_x,k_y) &=& a_y(-k_x+\pi,k_y+\pi)
\end{eqnarray} 
Similarly, the invariance under $G_{P_y}P_y$ leads to
\begin{eqnarray}
a_x(k_x,k_y) = a_x(k_x+\pi,-k_y+\pi),  \nonumber\\
a_y(k_x,k_y) = -a_y(k_x+\pi,-k_y+\pi)
\end{eqnarray}
The symmetry of $\mathbf{a}$ is summarized in Fig. \ref{fig2}a

For a loop $C$ formed by the two segment $C_1$ and $C_2$ in Fig.
\ref{fig2}a,  we find that $\int_{C_1} d\mathbf{k}\cdot \mathbf{a} =
\int_{C_2} d\mathbf{ k}\cdot \mathbf{a} $ and $2(2\pi)^{-1}\int_{C_1}
d\mathbf{k}\cdot \mathbf{a} $ is quantized as an integer.  Again, when
$2(2\pi)^{-1}\int_{C_1} d\mathbf{k}\cdot \mathbf{a} $ is an odd integer,
$\det(H(\mathbf{k}))$ must vanish at least at one point on the line
$((0,0),(\pi,\pi))$.

A typical distribution of the zeros of $\det(H(\mathbf{k}))$ may look like
Fig.  \ref{fig2}b.  The pattern of the winding number is determined by noting
that two points related by $G_{P_x}P_x$ or $G_{P_y}P_y$ have the opposite
winding numbers and two points related by $G_{P_{xy}}P_{xy}$ have the same
winding number (see Fig. \ref{fig2}b).  The quantum orders in the spin liquid
are again characterized by (at least) two universal properties: PSG and TOZ.
For Z2Azz13 spin liquids, the TOZ is again a collection of the following data:
(a) a pattern of $+1$ and $-1$ zeros along the line $((0,0),(\pi,\pi))$, (b)
$(N_+, N_-)$, the number of $(+1, -1)$ zeros \emph{inside} the triangle
$((0,0),(\pi,\pi), (0,\pi))$ (not including the zeros on the sides and
corners).

Now, let us consider the transitions between quantum orders with a fixed PSG
Z2Azz13.  Again, by examine the possible ways in which TOZ can change that are
consistent with the symmetry, we find that the quantum phase
transitions (caused by tuning one parameter) can have, at least, the follow 4
classes.  Because of the symmetry, we only need to describe the motion of zeros
near a triangle $((0,0), (\pi,0),(\pi,\pi))$.\\
(A) A type 1c transition on the line $((0,0), (\pi,\pi))$.  \\
(B) A type 2a transition on the line $((0,0), (\pi,\pi))$ .  \\
(C) A type 3 transition at $(0,\pi)$ .  \\
(D) A type 1a transition \emph{inside} the triangle $((0,0),(\pi/2,\pi/2),
(0,\pi/2))$.
%(A) A pair of $(+1,-1)$ zeros is created/annihilated on the line $((0,0),
%(\pi,\pi))$ at the phase transition.  The two zeros split in the direction
%along the line after creation or before annihilation.\\
%(B) Two $+1$ zeros merge along the diagonal line $((0,0),(\pi,\pi))$ at
%$\mathbf{k}=(0,0)$ (or $(\pi,\pi)$) and two $-1$ zeros marge along the other
%diagonal line at $(\pi,\pi)$ (or $\mathbf{k}=(0,0)$) to reach the transition
%point.  Two $-1$ zeros split along the diagonal $((0,0),(\pi,\pi))$ line at
%$\mathbf{k}=(\pi,\pi)$ (or $(0,0)$) and two $+1$ zeros split along the other
%diagonal line at $(0,0)$ (or $\mathbf{k}=(\pi,\pi)$) to move away from the
%transition point.\\ 
%(C) Two $+1$ zeros or two $-1$ zeros along the diagonal lines $((0,0),
%(\pi,\pi))$ merge and split perpendicular to the diagonal lines.\\
%(D) A pair of $(+1,-1)$ zeros is created/annihilated \emph{inside} the
%triangle $((0,0),(\pi/2,\pi/2), (0,\pi/2))$.

Since the motion of zeros are restricted by the symmetries of $H(\mathbf{k})$,
we find that those motion can only change a quantized winding number
$(2\pi)^{-1}\int_{C_3} d\mathbf{k}\cdot \mathbf{a} $ by an even integer,
where $C_3$ is the loop $(0,0) \to (0,\pi) \to (\pi,\pi) \to (0,0)$.  Thus
$(2\pi)^{-1}\int_{C_3} d\mathbf{k}\cdot \mathbf{a} $ mod $2$ is a quantum
number which is the same for all the Z2Azz13 ansatz. A direct calculation
shows that
$(2\pi)^{-1}\int_{C_1} d\mathbf{k}\cdot \mathbf{a} $ mod $2$ is equal to 1.
Therefore, there must be at least one zero on each of the four diagonal lines
$((0,0), (\pm \pi, \pm\pi))$.  The Z2Azz13 spin liquids at least have 4
two-component Dirac fermions. In general it has $4+8n$ two-component Dirac
fermions. See that PSG alone can some times guarantee the existence of gapless
fermions.

\subsection{Gapless fermions in Z2Bx2(12)n spin liquids}

The third $Z_2$ spin liquid that we are going to study is the Z2Bx2(12)n
spin liquid. A simple example of this class of ansatz is given by\cite{Wqo}
\begin{eqnarray}
\label{Z2Bx2(12)n}
u_{\mathbf{i},\mathbf{i}+\hat{\mathbf{x}}}&=&i\chi \tau^0  + \eta \tau^2 
 \nonumber\\
u_{\mathbf{i},\mathbf{i}+\hat{\mathbf{y}}}&=&(-)^{i_x}(i\chi\tau^0+\eta\tau^1)
 \nonumber\\
 a^{1,2,3}_0 &=& 0  .
\end{eqnarray}
The Z2Bx2(12)n PSG is generated by
\begin{eqnarray}
 G_x(\mathbf{i}) &=& (-)^{i_y} \tau^0, \ \ \ \ \  
G_y(\mathbf{i}) = \tau^0 ,  
\nonumber\\
 G_{P_x}(\mathbf{i})&=& i(-)^{i_x}\tau^1,\ \ \ \ \ 
G_{P_y}(\mathbf{i}) = i(-)^{i_y}\tau^2,
\nonumber\\
 G_{P_{xy}}(\mathbf{i})&=& i(-)^{i_xi_y}\tau^{12}, \ \ \ \ \ 
 G_T(\mathbf{i}) = (-)^{i_x+i_y}\tau^0, \nonumber\\
 G_0(\mathbf{i}) &=& (-)^{i_x+i_y}\tau^0, 
\end{eqnarray}
where $\tau^{ab}\equiv (\tau^a+\tau^b)/\sqrt{2}$.
The fermion spectrum for ansatz Eq.~\ref{Z2Bx2(12)n} is determined by
\begin{eqnarray}
 H &=& -2\chi \sin(k_x) \Gamma_0 + 2\eta \cos(k_x) \Gamma_2 \nonumber\\
   & & -2\chi \sin(k_y) \Gamma_1 + 2\eta \cos(k_y) \Gamma_3
\end{eqnarray}
where $k_x \in (-\pi/2,\pi/2)$, $k_y \in (-\pi/2,\pi/2)$ and
\begin{eqnarray}
 \Gamma_0 &=& \tau^0\otimes \tau^3\otimes \tau^0, \ \ \ \ \
 \Gamma_2 = \tau^2\otimes \tau^3\otimes \tau^0, \nonumber\\
 \Gamma_1 &=& \tau^0\otimes \tau^1\otimes \tau^3, \ \ \ \ \
 \Gamma_3 = \tau^1\otimes \tau^1\otimes \tau^3  .
\end{eqnarray}
$H$ acts on the following eight-component fermion field
$\Psi^T_{\mathbf{k}}=(\psi^T_{\mathbf{k}},  \psi^T_{\mathbf{k}+\pi \hat{\mathbf{x}}},
\psi^T_{\mathbf{k}+\pi\hat{\mathbf{y}}},  \psi^T_{\mathbf{k}+\pi \hat{\mathbf{x}}+\pi\hat{\mathbf{y}}})$.
We note that $H(\mathbf{k})$ is \emph{not} periodic in the 
Brillouin zone, $k_x\in (-\pi/2,\pi/2)$ and $k_y\in (-\pi/2,\pi/2)$.
It satisfies a twisted periodic boundary condition
\begin{eqnarray}
\label{bcon}
 B_x H(\mathbf{k}) B_x &=& H(\mathbf{k} + \pi\hat{\mathbf{x}}) \nonumber\\
 B_y H(\mathbf{k}) B_y &=& H(\mathbf{k} + \pi\hat{\mathbf{y}}) \nonumber\\
 B_x &=& \tau^0\otimes\tau^1\otimes\tau^0 \nonumber\\
 B_y &=& \tau^0\otimes\tau^0\otimes\tau^1 
\end{eqnarray}

The fermion spectrum can be calculated exactly and its four branches take a
form $\pm E_1(\mathbf{k})$ and $\pm E_2(\mathbf{k})$.  The fermion energy vanishes at
two isolated points $\mathbf{k}=(0,0), (0,\pi)$. Near $\mathbf{k} =0$ the low energy
spectrum is given by 
\begin{equation}
E=\pm \eta^{-1} \sqrt{(\chi^2+\eta^2)^2( k_x^2-k_y^2)^2 + 4 \chi^4 k_x^2k_y^2}
\end{equation}
It is interesting to see that the energy does not vanish linearly as $\mathbf{
k}\to 0$, instead it vanishes like $\mathbf{k}^2$.

Now let us consider the symmetry of a generic Z2Bx2(12)n Hamiltonian
$H(\mathbf{k})$ as required by the Z2Bx2(12)n PSG.  The invariance of the
ansatz under $G_xT_x$ leads to
\begin{eqnarray}
U_x H(\mathbf{k}) U_x  &=& H(\mathbf{k}),  \nonumber\\
U_x &=& \tau^0\otimes\tau^3\otimes\tau^1
\end{eqnarray}
and the invariance of the ansatz under $G_yT_y$ leads to
\begin{eqnarray}
U_y H(\mathbf{k}) U_y  &=& H(\mathbf{k}),  \nonumber\\
U_y &=&  \tau^0\otimes\tau^0\otimes \tau^3 
\end{eqnarray}
Therefore, $H(\mathbf{k})$ has a form
\begin{equation}
 H(\mathbf{k}) = 
\left(\begin{array}{ll}  h_{4\times 4}(\mathbf{k}) & 0 \\
      0 & h_{4\times 4}(\mathbf{k}) \end{array}\right)
\end{equation}

Under the $G_{P_x}P_x$ transformation, we have
\begin{eqnarray}
U_{P_x} H(k_x,k_y) U^\dagger_{P_x} &=& H(-k_x,k_y) \nonumber\\
U_{P_x} &=& \tau^1\otimes\tau^1\otimes\tau^0 
\end{eqnarray}
Under the $G_{P_y}P_y$ transformation, we have
\begin{eqnarray}
U_{P_y} H(k_x,k_y) U^\dagger_{P_y} &=& H(k_x,-k_y) \nonumber\\
U_{P_y} &=& \tau^2\otimes\tau^0\otimes\tau^1 
\end{eqnarray}

To calculate the $G_{P_{xy}}P_{xy}$ transformation,
we note that $(-)^{i_xi_y} =(1+(-)^{i_x}+(-)^{i_y}-(-)^{i_x+i_y})/2$ and
the gauge transformation $(-)^{i_xi_y}$ changes $\psi_{\mathbf{k}}$ to
\begin{equation}
 \frac{1}{2}(\psi_{\mathbf{k}} + \psi_{\mathbf{k}+\pi\hat{\mathbf{x}}}
+ \psi_{\mathbf{k}+\pi\hat{\mathbf{y}}} - \psi_{\mathbf{k}+\pi(\hat{\mathbf{x}}+\hat{\mathbf{y}})} ) 
\end{equation}
or
\begin{eqnarray}
 \Psi_{\mathbf{k}} 
\to &
\frac{1}{2} \left(
\tau^0\otimes\tau^0\otimes\tau^0 
+\tau^0\otimes\tau^1\otimes\tau^0  \right. \nonumber\\
&\ \ \ \ \ \left.+\tau^0\otimes\tau^0\otimes\tau^1
-\tau^0\otimes\tau^1\otimes\tau^1
\right) \Psi_{\mathbf{k}}  \nonumber\\
\end{eqnarray}
We also note that $P_{xy}$ switches
$\psi_{\mathbf{k}+\pi\hat{\mathbf{x}}}$ and $\psi_{\mathbf{k}+\pi\hat{\mathbf{y}}}$ in
$\Psi_{\mathbf{k}}$. Thus it changes 
\begin{eqnarray}
 \Psi_{k_x, k_y} 
\to &
%\frac{1}{4}\left(
%\tau^0\otimes(\tau^0+\tau^3)\otimes(\tau^0+\tau^3)\right. \nonumber\\
%&\ \ \ \ \ +
%\tau^0\otimes(\tau^0-\tau^3)\otimes(\tau^0-\tau^3) \nonumber\\
%&\ \ \ \ \ +
%\tau^0\otimes(\tau^1+i\tau^2)\otimes(\tau^1-i\tau^2) \nonumber\\
%&\ \ \ \ \ + \left.
%\tau^0\otimes(\tau^1-i\tau^2)\otimes(\tau^1+i\tau^2)
%\right)
% \Psi_{k_y, k_x} \nonumber\\
%&=& 
\frac{1}{2}\left(
\tau^0\otimes\tau^0\otimes\tau^0 +
\tau^0\otimes\tau^3\otimes\tau^3 \right. \nonumber\\
&\ \ \ \ \ \left. +
\tau^0\otimes\tau^1\otimes\tau^1 +
\tau^0\otimes\tau^2\otimes\tau^2
\right)
 \Psi_{k_y, k_x}
\end{eqnarray}
Therefore, the $G_{P_{xy}}P_{xy}$ transformation changes 
\begin{eqnarray}
 \Psi_{k_x, k_y} 
&\to &
\frac{1}{4} \left(
\tau^{12}\otimes\tau^0\otimes(\tau^0+\tau^1) 
+\tau^{12}\otimes\tau^1\otimes(\tau^0-\tau^1)
\right)   \nonumber\\
&&\ \ \left(
\tau^0\otimes\tau^0\otimes\tau^0 +
\tau^0\otimes\tau^3\otimes\tau^3 \right. \nonumber \\
&&\ \ \ \ \ +\left.
\tau^0\otimes\tau^1\otimes\tau^1 +
\tau^0\otimes\tau^2\otimes\tau^2
\right)
 \Psi_{k_y, k_x}  \nonumber \\
&= &
\frac{1}{2} \left(
\tau^{12}\otimes\tau^2\otimes\tau^2+
\tau^{12}\otimes\tau^3\otimes\tau^3 \right. \nonumber\\
&&\  \left. +
\tau^{12}\otimes\tau^1\otimes\tau^0+
\tau^{12}\otimes\tau^0\otimes\tau^1
\right)
 \Psi_{k_y, k_x}   \nonumber\\
&=& U_{P_{xy}} \Psi_{k_y, k_x}
\end{eqnarray}
One can check that $(U_{P_{xy}})^2 =1$.  The $G_{P_{xy}}P_{xy}$ invariance
implies that
\begin{equation}
U_{P_{xy}} H(k_x,k_y) U^\dagger_{P_{xy}} =  
 H(k_y,k_x)
\end{equation}

Next, we study the symmetry of $\mathbf{a}$ in order to understand gapless
fermions for a generic Z2Bx2(12)n ansatz.  First we consider the
transformation $ G_TT$.  We know that $u_{\mathbf{i}\mathbf{j}}$ and hence
$H(\mathbf{k})$ change sign under $G_TT$. Since the action $(-)^{i_x+i_y}$
changes $\Psi_{\mathbf{k}} \to \tau^0\otimes \tau^1\otimes\tau^1
\Psi_{\mathbf{k}}$, we find $\gamma^5_{00} = \tau^0\otimes\tau^1\otimes\tau^1$
anticommutes with $H(\mathbf{k})$ for any Z2Bx2(12)n ansatz.  Since $U_x$ and
$U_y$ commute with the Hamiltonian $H(\mathbf{ k})$ we can construct four
different $\gamma^5$'s that anticommute with $H(\mathbf{k})$.  The  four
$\gamma^5$'s are given by
\begin{equation}
 \gamma^5_{mn} = i(i)^{(m-1)(n-1)}\gamma^5_{00} U_x^m U_y^n, \ \ \ \ \ m,n = 0,1
\end{equation}
We can use $\gamma^5_{mn}$ to define four $\mathbf{a}^{mn}$ and four
topologically invariant winding numbers.  We note that
$\{\gamma^5_{00},U_x\}=0$, $\{ \gamma^5_{00},U_y\}=0$ and $\{ U_x,U_y\}=0$.
Thus, $\gamma^5_{00}$, $\gamma^5_{10}$ and $\gamma^5_{10}$ anticommute with at
least one of the $U_x$ and $U_y$.  Since $U_x$ and $U_y$ commute with
$H(\mathbf{ k})$ and hence $M(\mathbf{k})$, we can show that the winding
numbers from $\gamma^5_{00}$, $\gamma^5_{10}$ and $\gamma^5_{01}$ are always
zero. Since $U_x$ and $U_y$ anticommute, we can choose a basis such that $U_x
= \tau^0\otimes\tau^0\otimes\tau^1$ and $U_y =
\tau^0\otimes\tau^0\otimes\tau^3$. In this basis $H(\mathbf{k})$ and
$\gamma^5_{11}$ have a form $M_{4\times 4}\otimes \tau^0$ since they commute
with both $U_x$ and $U_y$. Therefore, the winding numbers from $\gamma^5_{11}$
are always even integers.  In the following we will only consider the
$\gamma^5_{11}$-winding number.  We will set 
\begin{equation}
\gamma^5=\gamma^5_{11}= \tau^0\otimes\tau^2\otimes\tau^3
\end{equation}
and $\mathbf{a} = \mathbf{a}^{11}$.

From the Z2Bx2(12)n PSG, we see that $G_TG_xT_xG_yT_y$ and $G_{P_x}P_x$
anticommute.  Therefore, $\gamma^5$ and $U_{P_x}$ anticommute and we find that
the $G_{P_x}P_x$ invariance requires
\begin{eqnarray}
a_x(k_x,k_y) &=& -(-a_x(-k_x,k_y)),  \nonumber\\
a_y(k_x,k_y) &=& -(a_y(-k_x,k_y)) .
\end{eqnarray}
Similarly, we also find that the $G_{P_y}P_y$ invariance requires 
\begin{eqnarray}
a_x(k_x,k_y) &=& -(a_x(k_x,-k_y)), \nonumber\\
a_y(k_x,k_y) &=& -(-a_y(k_x,-k_y)) .
\end{eqnarray}
From $U_{P_{xy}} U_x  U_{P_{xy}} = U_y$, $U_{P_{xy}} \gamma^5_{00}  U_{P_{xy}} =
\gamma^5_{00}$, and
\begin{equation}
 U_{P_{xy}} \gamma^5_{mn}  U_{P_{xy}} = \gamma^5_{nm}(-)^{mn},
\end{equation}
we find that the $G_{P_{xy}}P_{xy}$ invariance requires 
\begin{eqnarray}
a_x(k_x,k_y) &=& -a_y(k_y,k_x) , \nonumber\\
a_y(k_x,k_y) &=& -a_x(k_y,k_x).
\end{eqnarray}
Due to the twisted boundary condition Eq.~\ref{bcon},
$\mathbf{a}$ is not a periodic function in the Brillouin zone, $k_x\in
(-\pi/2,\pi/2)$ and $k_y\in (-\pi/2,\pi/2)$.
Since $B_x$ and $B_y$ anticommutes with $\gamma^5$, we find
\begin{eqnarray}
 \mathbf{a}(\mathbf{k}) = - \mathbf{a}(\mathbf{k}+\pi\hat{\mathbf{x}}),\ \ \ \
\
 \mathbf{a}(\mathbf{k}) = - \mathbf{a}(\mathbf{k}+\pi\hat{\mathbf{y}}).
\end{eqnarray}
This allows us to obtain distribution of zeros and winding numbers over an
extended Brillouin zone.

\begin{figure}
\centerline{
\includegraphics[height=1.5in]{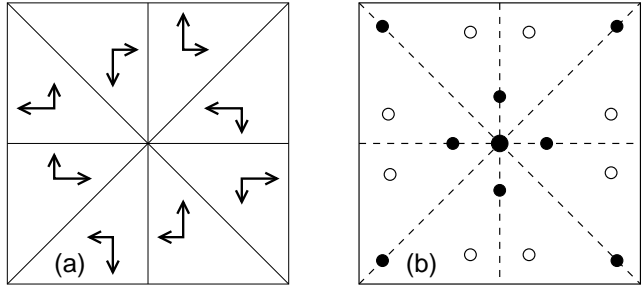}
}
\caption{
(a) The symmetry of $\mathbf{a}$ in the Brillouin zone, $k_x\in
(-\pi/2,\pi/2)$ and $k_y\in (-\pi/2,\pi/2)$, for the Z2Bx2(12)n state.  The
two arrows represent $a_x \hat{\mathbf{x}}$ and $a_y \hat{\mathbf{y}}$.  
(b) A typical distribution of the zeros of $\det(H(\mathbf{k}))$. A solid dot represents a
zero with winding number $+2$ and an empty dot represents a
zero with winding number $-2$. 
The large dot at $\mathbf{k}=(0,0)$ has a winding number 4.
}
\label{fig3}
\end{figure}

From the symmetry of $\mathbf{a}$, we find a typical distribution of the zero
of $\det(H(\mathbf{k}))$ may looks like Fig.  \ref{fig3}b.  The pattern of the
winding number is determined by noting that two points related by
$G_{P_x}P_x$, $G_{P_y}P_y$, or $G_{P_{xy}}P_{xy}$ have the same winding
number. The quantum orders in the spin liquid are again characterized by (at
least) two universal properties: PSG and TOZ.  For Z2Bx2(12)n spin liquids,
the TOZ is a collection of the following data: (a) $(N_+, N_-)$, the number of
$(+2, -2)$ zeros \emph{inside} the triangle $((0,0),(\pi/2,\pi/2), (0,\pi/2))$
(not including the zeros on the sides and corners).  (b) a pattern of $+2$ and
$-2$ zeros on the diagonal line $((0,0),(\pi/2,\pi/2))$.  

The total sum of the winding numbers in the Brillouin zone is always zero for
the Z2A0013 and Z2Azz13 spin liquids, since their $H(\mathbf{k})$ is a
periodic function in the Brillouin zone. For the Z2Bx2(12)n spin liquid,
$H(\mathbf{k})$ satisfies a twisted boundary condition Eq.~\ref{bcon}.  The
total sum of the winding numbers $N_{tot}$, in this case, can be non-zero.
From the symmetry of $\mathbf{a}$, we see that $N_{tot}$ mod 8 is a
topological number that is the same for any two ansatz that can be
continuously deformed into each other via a path that does not change the
Z2Bx2(12)n PSG.  For the Z2Bx2(12)n spin liquid, the $N_{tot}$ mod 8 is found
to be $4$. Therefore, $\det(H(\mathbf{ k}))$ always has a zero at
$\mathbf{k}=(0,0)$ with a winding number $4+8n$. A generic zero at an
arbitrary $\mathbf{k}$ has a winding number $\pm 2$ which gives rise to two
two-component Dirac fermions with a linear dispersion $E(\mathbf{k}) \propto
|\mathbf{k}|$.  The $\pm 4$ winding number for the zero at $\mathbf{k}=(0,0)$
implies that there are two gapless two-component ``Dirac'' fermions with a
quadratic dispersion relation $E(\mathbf{k}) \propto (\mathbf{k})^2$. We see
that quadratic dispersion relation at $\mathbf{k} =(0,0)$ is protected by the
Z2Bx2(12)n PSG and is a generic property of the Z2Bx2(12)n spin liquids.

Now let us consider the transitions between quantum orders with a fixed PSG
Z2Bx2(12)n.  
%By examine the possible ways in which TOZ can change that are
%consistent with the symmetry, we find that the possible quantum phase
%transitions (caused by tuning one parameter) can be divided into the follow 6
%classes. Again, we only need to describe the motion of zeros near a triangle
%$((0,0), (\pi/2,0),(\pi/2,\pi/2))$.\\ 
We find, at least, the following 6 classes of transitions.\\
(A) A type 1b transition on the side $((0,0), (\pi/2,0))$ or side
$((\pi/2,0),(\pi/2,\pi/2))$ of the triangle.  Note the winding number changes
sign if we shift $\mathbf{k}$ by $\pi \hat{\mathbf{x}}/2$ or $\pi
\hat{\mathbf{y}}/2$.\\ 
(B) A type 1c transition on the side $((0,0),(\pi/2,\pi/2))$ of the triangle.
\\ 
(C) A type 4 transition at $\mathbf{k}=(0,0)$. \\ 
(D) A type 3 transition (in the extended Brillouin zone) at $\mathbf{
k}=(\pi/2,\pi/2)$.  Note the winding number changes sign if we shift
$\mathbf{k}$ by $\pi \hat{\mathbf{x}}/2$ or $\pi \hat{\mathbf{y}}/2$.\\ 
(E) A type 2a transition on the side $((0,0),(\pi/2,\pi/2))$ of the triangle.
\\ 
(F) A type 1a transition \emph{inside} the triangle $((0,0),(\pi/2,0),
(\pi/2,\pi/2))$.

\subsection{Gapless fermions in the U1Cn01n spin liquids}

Last, we would like to study gapless fermions in the U1Cn01n spin liquid (the
staggered-flux/$d$-wave phase of the slave-boson theory
\cite{AM8874,KL8842}).  A simple ansatz for
the U1Cn01n state is given by 
\begin{eqnarray} a_0^l&=&
0\nonumber\\
u_{\mathbf{i},\mathbf{i}+\hat{\mathbf{x}}} &=& \chi \tau^1 +\eta \tau^2  \nonumber\\
u_{\mathbf{i},\mathbf{i}+\hat{\mathbf{y}}} &=& \chi \tau^1 -\eta \tau^2
\end{eqnarray}
More general ansatz can be found in Ref. \cite{Wqo}.
The U1Cn01n PSG is generated by
\begin{eqnarray}
 G_x(\mathbf{i}) &=& \tau^0,\ \ \ \ \ G_y(\mathbf{i}) = \tau^0,  \nonumber\\
 G_{P_x}(\mathbf{i}) &=& \tau^0,\ \ \ \ \ G_{P_y}(\mathbf{i}) = \tau^0,  
\nonumber\\
 G_{P_{xy}}(\mathbf{i}) &=& i\tau^1,\ \ \ \ \ G_T(\mathbf{i}) = i\tau^3, 
\nonumber\\
 G_0(\mathbf{i}) &=& e^{i(-)^{i_x+i_y}\theta \tau^3}, 
\end{eqnarray}

\begin{figure}
\centerline{
\includegraphics[height=1.5in]{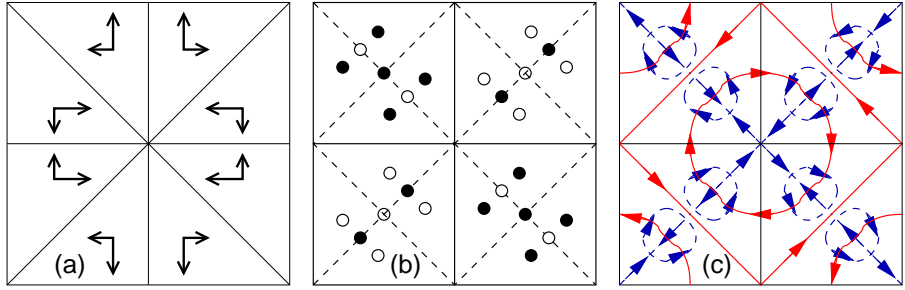}
}
\caption{
(a) The symmetry of $\mathbf{a}$ in the Brillouin zone. The two arrows represent
$a_x \hat{\mathbf{x}}$ and $a_y \hat{\mathbf{y}}$.
(b) A typical distribution of the zeros of $\det(H(\mathbf{k}))$. A solid dot represents a
zero with winding number $+1$ and an empty dot represents a
zero with winding number $-1$.
(c) The curves $\epsilon=0$ (solid lines) and $\Delta = 0$ (dash lines). 
%The direction of the curves allows us to define the righ-hand-side of the curve
%where $\epsilon$ (or $\Delta$) is positive.
}
\label{fig4}
\end{figure}

Since the Hamiltonian $u_{\mathbf{i}\mathbf{j}}$, $G_xT_x$ and $G_yT_y$ commute, the $\mathbf{
k}$-space Hamiltonian $H(\mathbf{k})$ can be defined which is periodic over the
Brillouin zone, $k_x\in (-\pi,\pi)$ and $k_y\in (-\pi,\pi)$.  Since
$G_T$ and the Hamiltonian $u_{\mathbf{i}\mathbf{j}}$ anticommute, we can choose
$\gamma^5 = -iG_T$ and use such a $\gamma^5$ to define $\mathbf{a}$. Now we can find the
symmetry of $\mathbf{a}$ through the algebraic property of the U1Cn01n PSG.
We first note that $G_T$ commute with both $G_{P_x}P_x$ and $G_{P_y}P_y$.
Thus, the $G_{P_x}P_x$ and $G_{P_y}P_y$ invariance of the ansatz implies
\begin{eqnarray}
 a_x(k_x,k_y) &=& -a_x(-k_x,k_y), \ \ \ \ \
 a_y(k_x,k_y) = a_y(-k_x,k_y); \nonumber\\ 
 a_x(k_x,k_y) &=& a_x(k_x,-k_y), \ \ \ \ \
 a_y(k_x,k_y) = -a_y(k_x,-k_y).
\end{eqnarray}
Since $G_{P_{xy}}P_{xy}$ anticommutes with $G_T$, we have, under
$G_{P_{xy}}P_{xy}$, 
\begin{equation}
 a_x(k_x,k_y) = -a_y(k_y,k_x), \ \ \ \ \ 
 a_y(k_x,k_y) = -a_x(k_y,k_x).
\end{equation}
From $[G_0, H]=0$, we find $L_0\equiv (-)^{i_x+i_y}\tau^3$ commute with the
Hamiltonian $u_{\mathbf{i}\mathbf{j}}$. Since $L_0$ and $G_T$ commute and the action of
$L_0$ increase $\mathbf{k}$ by $(\pi,\pi)$, we have
\begin{eqnarray}
 a_x(k_x,k_y) &=& a_x(k_x+\pi,k_y+\pi),  \ \ \ \ \
 a_y(k_x,k_y) = a_y(k_x+\pi,k_y+\pi).
\end{eqnarray}
The symmetry of $\mathbf{a}$ is summarized in Fig. \ref{fig4}a.  Comparing
with the Z2A0013 state, here we have an additional symmetry $\mathbf{k} \to
\mathbf{k} +\pi\hat{\mathbf{x}}+\pi\hat{\mathbf{y}}$ caused by $L_0$.
The $\mathbf{k}$ Hamiltonian $H(\mathbf{k})$ still has the form Eq.
\ref{Hk0013}.

From the symmetry of $\mathbf{a}$, we find that a typical distribution of the
zero of $\det(H(\mathbf{k}))$ may looks like Fig.  \ref{fig4}b.  The pattern
of the winding number is determined by noting that two zeros related by
$G_{P_x}P_x$ or $G_{P_y}P_y$ have the opposite winding number and two zeros
related by $G_{P_{xy}}P_{xy}$ have the same winding number.  Two zeros related
$L_0$ (a shift of $(\pi,\pi)$ in $\mathbf{k}$) also have the same winding
number.

The quantum orders in the spin liquid are characterized by (at least) two
universal properties: PSG and TOZ.  For U1Cn01n spin liquids, the TOZ is a
collection of the following data: (a) $(N_+, N_-)$, the number of $(+1, -1)$
zeros \emph{inside} the triangle $((0,0),(\pi/2,\pi/2), (0,\pi))$ (not
including the zeros on the sides and corners).  (b) a pattern of $+1$ and $-1$
zeros on the two sides of the triangle $((0,0),(\pi/2,\pi/2))$ and
$((0,\pi),(\pi/2,\pi/2))$.  

Let $N_{tot}$ be the total sum of the winding numbers in the square $((0,0),
(0,\pi), (\pi,\pi), (\pi,0))$, which is a quarter of the Brillouin zone.  From
the distribution and the restricted motion of the zeros, we see that $N_{tot}$
mod 2 is a topological number that is the same for any two ansatz that can be
continuously deformed into each other via a path that does not change the
U1Cn01n PSG.  For the U1Cn01n spin liquid, the $N_{tot}$ mod 2 is found to be
$1$.  Therefore, $\det(H(\mathbf{k}))$ always has a zero at $\mathbf{
k}=(\pi/2,\pi/2)$ with a winding number $1+2n$.  We see that the zero at
$(\pi/2,\pi/2)$ is protected by the U1Cn01n PSG and a generic U1Cn01n spin
liquids have at least four two-component Dirac fermions at $\mathbf{
k}=(\pm\pi/2,\pm\pi/2)$.

Now let us consider the transitions between quantum orders with a fixed PSG
U1Cn01n.  We find that the possible quantum phase transitions (caused by
tuning one parameter) can be, at least, divided into the follow 6 classes. \\
%We only need to describe the motion of zeros near a triangle
%$((0,0), (\pi/2,\pi/2) , (0,\pi))$.\\
(A) A type 1b transition on the side $((0,0), (0,\pi))$ of the triangle.\\
(B) A type 1c transition on the side $((0,0),(\pi/2,\pi/2))$ or side
$((0,\pi),(\pi/2,\pi/2))$ of the triangle.\\
(C) A type 3 transition at $\mathbf{k}=(0,0)$. \\
(D) A type 2b transition at $\mathbf{k}=(\pi/2,\pi/2)$ which changes the winding
number of the zero at $(\pi/2,\pi/2)$ from 1 to $-1$ (or from $-1$ to $1$).\\
(E) A type 2a transition on the side $((0,0),(\pi/2,\pi/2))$ or the side
$((0,\pi),(\pi/2,\pi/2))$ of the riangle.\\
(F) A type 1a transition \emph{inside} the triangle $((0,0),(\pi/2,0),
(\pi/2,\pi/2))$.

The analysis in this subsection also illustrates that the gapless fermions and
the TOZ are directly determined from the algebra of the PSG. We can obtain
those results even without writing down the explicit forms of $H(\mathbf{k})$,
$G_T$, $G_{P_x}P_x$, {\it etc.}.

\section{Summary}

In this paper, we studied the relation between quantum orders and gapless
fermion excitations in a quantum spin-1/2 system on a 2D square lattice. By
studying the symmetry of $H(\mathbf{k})$ as imposed by the PSG and using the
winding number, we find that PSG directly determine the pattern of Fermi
points in the Brillouin zone. Thus the gapless fermion excitations (the
spinons) are originated from and protected by the quantum orders.

We would like to thank H.T. Nieh and Z.Y. Weng for warm hospitality in Center
for Advanced Study, Tsinghua University, where this work was started.  XGW is
supported by NSF Grant No.  DMR--01--3156 and by NSF-MRSEC Grant No.
DMR--98--08941.

\end{document}